\documentclass[letterpaper, 10 pt, conference]{ieeeconf}

\IEEEoverridecommandlockouts  
\usepackage{amsfonts}
\usepackage{amsmath}
\usepackage[dvipsnames]{xcolor}
\newtheorem{theorem}{Theorem}

\usepackage{algorithm,algorithmic}
\usepackage{graphicx}
\graphicspath{ {./figs/} }
\usepackage{caption}
\usepackage{subcaption}

\newtheorem{definition}{Definition}
\newtheorem{lemma}{Lemma}

\usepackage{amsmath,amssymb}

\newcommand{\X}{\mathcal{X}}

\newcommand{\F}{\mathcal{F}}

\title{\LARGE \bf Robust Model Predictive Shielding for Safe Reinforcement Learning with Stochastic Dynamics}

\author{Shuo Li$^{1}$ and Osbert Bastani$^{2}$
\thanks{$^{1}$Shuo Li is with the GRASP Lab, University of Pennsylvania, USA
        {\tt\small lishuo1@seas.upenn.edu}}%
\thanks{$^{2}$Osbert Bastani is with the Department of Computer and Information Science, University of Pennsylvania, USA
        {\tt\small obastani@seas.upenn.edu}}%
}

\begin{document}

\maketitle
\thispagestyle{empty}
\pagestyle{empty}

\begin{abstract}
This paper proposes a framework for safe reinforcement learning that can handle stochastic nonlinear dynamical systems. We focus on the setting where the nominal dynamics are known, and are subject to additive stochastic disturbances with known distribution. Our goal is to ensure the safety of a control policy trained using reinforcement learning, e.g., in a simulated environment. We build on the idea of model predictive shielding (MPS), where a backup controller is used to override the learned policy as needed to ensure safety. The key challenge is how to compute a backup policy in the context of stochastic dynamics. We propose to use a tube-based robust NMPC controller as the backup controller. We estimate the tubes using sampled trajectories, leveraging ideas from statistical learning theory to obtain high-probability guarantees. We empirically demonstrate that our approach can ensure safety in stochastic systems, including cart-pole and a non-holonomic particle with random obstacles.
\end{abstract}

\section{Introduction}

There has been much recent progress in reinforcement learning (RL)~\cite{DBLP:journals/corr/HasseltGS15,DBLP:journals/corr/SchulmanLMJA15,DBLP:journals/corr/SchulmanWDRK17,DBLP:journals/corr/MnihBMGLHSK16,44806}.
As a consequence, there has been interest in using RL to design control policies for solving complex robotics tasks~\cite{DBLP:journals/corr/abs-1811-12927,DBLP:journals/corr/abs-1901-03737,DBLP:journals/corr/BojarskiTDFFGJM16}, including grasping~\cite{andrychowicz2018learning} and multi-agent planning~\cite{DBLP:journals/corr/abs-1907-05300,khan2019graph,tolstaya2019learning}. In particular, learning-enabled controllers (LECs) have the potential to outperform optimization-based controllers~\cite{article}. In addition, optimization-based controllers can often only be used under strong assumptions about the system dynamics, system constraints, and objective functions \cite{Garcia1989ModelPC,book}, which limits their applicability to complex robotics tasks.

However, safety concerns prevent LECs from being widely used in real-world tasks, which are often safety-critical in nature. In particular, unlike optimization-based controllers, it is typically infeasible to impose hard safety constraints on LECs. For example, even a simple safety property such as having a robot avoid static obstacles is nontrivial to impose. More complex safety properties, such as ensuring that a cart-pole or walking robot does not fall over, pose additional difficulty. Additionally, the real world typically has stochastic disturbances such as friction or wind, or the model used to simulate the dynamics may have errors; if an LEC is not robust to these perturbations, then it may fail catastrophically~\cite{DBLP:journals/corr/HoE16}. The issue is that the LEC is typically a black box machine learning model such as a deep neural network (DNN), and reasoning about the closed-loop behavior of the DNN poses computational challenges.

As a consequence, safe reinforcement learning has become an increasingly important area of research~\cite{gillula2012guaranteed,akametalu2014reachability,berkenkamp2017safe,bastani2018verifiable,alshiekh2018safe,fisac2019bridging,ivanov2019verisig}. Many methods in this area leverage optimization tools to prove that a learned neural network policy satisfies a given safety constraint~\cite{DBLP:journals/corr/abs-1903.01287,DBLP:journals/corr/abs-1906-04893,inproceedings,DBLP:journals/corr/abs-1802-03557,berkenkamp2017safe,bastani2018verifiable,ivanov2019verisig}. A related approach is \emph{shielding}, which verifies a simpler \emph{backup controller}, and then overrides the LEC using the backup controller when it can no longer ensure that using the LEC is safe~\cite{gillula2012guaranteed,akametalu2014reachability,alshiekh2018safe,DBLP:journals/corr/abs-1905-10691}. While these methods provide strong mathematical guarantees, they suffer from a number of shortcomings. For example, many of these methods scale exponentially in the state dimension, so they do not scale well to high-dimensional systems. Those that do typically rely on overapproximating reachable set of states, which can become very imprecise.

We build on a recently proposed idea called \emph{model predictive shielding} (MPS), which has been used to ensure safety of learned control policies~\cite{DBLP:journals/corr/abs-1803-08552,DBLP:journals/corr/abs-1905-10691}, including extensions to the multi-agent setting~\cite{zhang2019safe}. The basic idea is that rather than checking which states are safe ahead-of-time, we can dynamically check whether we can maintain safety if we use the LEC, and only use the LEC if we can do so. However, existing approach are limited---there has been work considering nonlinear dynamics, but assuming they are deterministic~\cite{DBLP:journals/corr/abs-1905-10691,zhang2019safe}, and there has been work considering stochastic or nondeterministic dynamics, but assuming they are linear~\cite{DBLP:journals/corr/abs-1803-08552}. Nonlinearity is important because many tasks where LECs have the most promise are highly nonlinear. Stochasticity is important for a number of reasons. For instance, there are often small perturbations in real-world dynamical systems. Similarly, it can be used to model estimation error in the robot's state (e.g., uncertainty in its position). Finally, LECs are often learned in simulation using a model of the dynamics; there are often errors in the model that need to be robustly accounted for.

We propose an approach, called robust MPS (RMPS), that extends MPS to the setting of stochastic nonlinear dynamics. Our approach uses robust nonlinear model-predictive control (NMPC) as the backup controller. The reason for using NMPC is that the goals of the backup controller are qualitatively different from the goals of the LEC. For example, consider the problem of building a robot that can run. The LEC tries to make the robot run as fast as possible. It may be able to outperform the robust NMPC, since the robust NMPC treats the stochastic perturbations conservatively. However, because RL lacks theoretical guarantees, the LEC cannot guarantee safety. Thus, we want to use the LEC as often as possible, but override it using a backup controller if we are not sure whether it is safe to use the LEC. The NMPC is an effective choice for the backup controller, where the goal is to stop the system and bring it to an equilibrium point, after which a feedback controller can be used to stabilize it. Continuing our example, the NMPC might bring the robot to a halt (e.g., where it is standing).


To achieve our goals, we build on algorithms for robust NMPC~\cite{doi:10.1080/00423114.2014.902537,6160389}, which aims to control a dynamical system in the presence of stochastic or nondeterministic perturbations, and on sampling-based estimation of forward reachable sets~\cite{allen2014machine,reist2016feedback}; in particular, we build closely on tube-based robust NMPC where sampling is used to estimate a tube within which the NMPC is guaranteed to stay (i.e., the tube is the forward reachable set of the NMPC)~\cite{doi:10.1002/rnc.1758,mayne2016robust}. However, these approaches do not provide any finite sample probabilistic guarantees on their estimates of the tubes.

We propose to use results from statistical learning theory to obtain provable probabilistic guarantees on our estimates of the sizes of the tubes~\cite{vapnik2013nature}. We develop a practical algorithm based on these theoretical results.
There has been recent work that provides theoretical guarantees on the estimated forward reachable set~\cite{liebenwein2018sampling}; however, they study the problem of obtaining \emph{underapproximations} of the forward reachable set, as opposed to \emph{overapproximations} that are needed for checking safety. Thus, their guarantees require qualitatively different techniques than the ones we use. To the best of our knowledge, we are the first to use these to construct estimates for forward reachable sets of nonlinear dynamical systems that come with probabilistic guarantees.


\textbf{Contributions.}
Our key contributions are: (i) an extension of the MPS algorithm to stochastic nonlinear dynamical systems (Section~\ref{sec:alg}), (ii) a novel algorithm for estimating tubes for RMPC with high-probability guarantees (Section~\ref{sec:alg}), and, as well as heuristic modifications that enable us to scale this algorithm to real-world systems, and (iii) experiments demonstrating how our approach ensures safety for LECs for cart-pole and for a single particle with non-holonomic dynamics and random obstacles (Section~\ref{sec:Exp}).

\section{Preliminaries}

\textbf{Dynamics.}
We consider stochastic nonlinear dynamics 
\begin{align*}
    x(k+1) = f(x(k), u(k)) + w(k),
\end{align*}
where $k$ is the time step, $x(k) \in \mathcal{X} \subseteq \mathbb{R}^{n_X}$ is the state, $u(k) \in \mathcal{U} \subseteq \mathbb{R}^{n_U}$ is the control input, and $w(k) \in \mathcal{W} \subseteq \mathbb{R}^{n_X}$ is a zero-mean stochastic perturbation with distribution $\mathcal{P}_{\mathcal{W}}$.

\textbf{Policy.}
A \emph{policy} is a map $\pi:\mathcal{X}\to\mathcal{U}$. A trajectory generated using $\pi$ from initial state $x\in\mathcal{X}$ is $\mathbf{x}=(x(0),x(1),...)$, where $x(0)=x$ and $x(k+1)=f^{(\pi)}(x(k))+w(k)$ and where $f^{(\pi)}(x)=f(x,\pi(x))$. Since $w(k)$ is random, $\mathbf{x}$ is a random sequence; we use $p(\mathbf{x}\mid\pi,x)$ to denote the probability of $\mathbf{x}$.

\textbf{Objective.}
We consider a cost function $\ell:\mathcal{X}\times\mathcal{U}\to\mathbb{R}$ and a discount factor $\gamma\in(0,1)$. Given a distribution $p(x)$ over initial states, the cost of a policy $\pi$ is
\begin{align*}
\ell(\pi) = \mathbb{E}_{p(x),p(\mathbf{x}\mid\pi,x)}\left[\sum_{k=0}^{\infty} \gamma^k \ell(x(k), u(k))\right].
\end{align*}

\textbf{Safety constraint.}
We consider a set of safe states $\mathcal{X}_{\text{safe}}\subseteq\mathcal{X}$, with the goal of ensuring that the system stays in $\mathcal{X}_{\text{safe}}$.
A trajectory $\mathbf{x}$ is safe if $x(k)\in\mathcal{X}_{\text{safe}}$ for all $k\ge0$.

\textbf{Shielding problem.}
Our goal is to construct a policy $\pi$ that achieves low cost while ensuring safety. In general, since the dynamics are stochastic, it is impossible to guarantee safety. Instead, our goal is to ensure that safety holds with high probability. We establish a theoretical safety guarantee in Theorem~\ref{thm:main}; we interpret this theorem  in Section~\ref{sec:rmpsdetails}.

Our approach is based on \emph{shielding}~\cite{gillula2012guaranteed,akametalu2014reachability,alshiekh2018safe,DBLP:journals/corr/abs-1905-10691}. This approach takes as input a policy $\hat{\pi}$ that optimizes $\ell(\pi)$, but does not ensure safety (though a penalty for violating safety may be encoded in $\ell$). We call $\hat{\pi}$ the \emph{learned policy}, since our key motivation is when $\hat{\pi}$ is a neural network policy trained using RL---e.g., we use DDPG~\cite{Silver:2014:DPG:3044805.3044850}. Nevertheless, our approach can be used with any controller $\hat{\pi}$.

Then, the \emph{shielding problem} is to construct a policy $\pi_{\text{shield}}$ that overrides $\hat{\pi}$ as needed to ensure safety. The key challenges are (i) determining how to override $\hat{\pi}$, and (ii) minimizing how often $\hat{\pi}$ is overridden.

\textbf{Notation.}
%
For $k\in\mathbb{N}$, we let $[k]=\{1,...,k\}$. We let $\mathbb{S}^n$ be the positive semidefinite matrices of dimension $n$. For $x\in\mathbb{R}^n$ and $M\in\mathbb{S}^n$, we let $\|x\|_M=x^{\top}Mx$. For $\mathcal{A}, \mathcal{B} \subseteq \mathbb{R}^n$,
\begin{align*}
\mathcal{A} \oplus \mathcal{B} &= \{\mathbf{a}+\mathbf{b}\mid \mathbf{a}\in\mathcal{A},~\mathbf{b}\in\mathcal{B}\} \\
\mathcal{A}\ominus\mathcal{B} &= \{\mathbf{c}\mid\{\mathbf{c}\}\oplus\mathcal{B}\subseteq\mathcal{A}\}
\end{align*}
denote their Minkowski sum and difference, respectively.

%


\begin{figure}[t]
\centering
\includegraphics[scale=0.25]{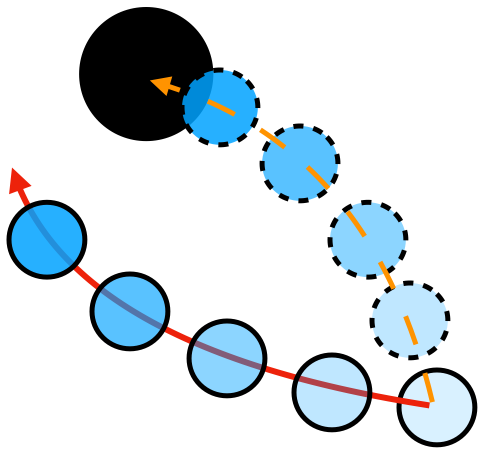}
\caption{An illustration of MPS. If using $\hat\pi$ (dashed orange line), the robot (blue) may unsafely crash into the obstacle (black). Using $\pi_{\text{backup}}$ (solid red line) ensures safety.}
\label{fig:MPS}
\end{figure}

\section{Background on Model Predictive Shielding}

Model predictive shielding (MPS) is a shielding algorithm that dynamically checks when to override $\hat{\pi}$. The key idea is to maintain an invariant that it can always use a \emph{recovery policy} $\pi_{\text{rec}}$ to safely transition to an equilibrium point~\cite{DBLP:journals/corr/abs-1803-08552,DBLP:journals/corr/abs-1905-10691,zhang2019safe}; a state $x\in\mathcal{X}$ that satisfies this invariant is \emph{recoverable} (denoted $x\in\mathcal{X}_{\text{rec}}$). Near the equilibrium point, we assume a linear feedback controller $\pi_{\text{stable}}$ can be used to ensure safety for an infinite horizon. Thus, as long as the system remains in $\mathcal{X}_{\text{rec}}$, then MPS can guarantee safety. The combination of $\pi_{\text{rec}}$ and $\pi_{\text{stable}}$ is the \emph{backup controller} $\pi_{\text{backup}}$ used to override $\hat{\pi}$. This approach is shown in Fig.~\ref{fig:MPS}.

For example, consider a driving robot. In this setting, $x$ is recoverable if the robot can safely apply the brakes to come to a stop. If $x$ is recoverable, but $f^{(\hat{\pi})}(x)$ is not, then MPS uses $\pi_{\text{backup}}$. Since $x$ is recoverable, using $\pi_{\text{rec}}$ is guaranteed to keep the system safe. Thus, MPS ensures safety for an infinite horizon from any recoverable state.

\begin{algorithm}[t]
\caption{Compute the RMPS controller for state $x$.}
\label{alg:shield}
\begin{algorithmic}
\renewcommand{\algorithmicrequire}{\textbf{procedure}}
\REQUIRE RMPS($x$) 
\IF{IsRecoverable($f^{(\hat{\pi})}(x)$)}
\STATE $\pi_{\text{backup}}\gets\varnothing$
\RETURN $\hat \pi(x)$
\ELSIF{$\pi_{\text{backup}}\neq\varnothing\wedge \text{IsRecoverable}(f^{(\pi_{\text{backup}})}(x)$)}
\RETURN $\pi_{\text{backup}}(x)$
\ELSE
\STATE $\pi_{\text{backup}}\gets\text{InitializeBackup}(x)$
\RETURN $\pi_{\text{backup}}(x)$
\ENDIF
\renewcommand{\algorithmicensure}{\textbf{Output:}}
\end{algorithmic} 
\end{algorithm} 

\begin{algorithm}[t]
\caption{Compute the backup controller for state $x$. It keeps internal state $(z_e,\mathcal{G}_e,\mathbf{x}^*,\mathbf{u}^*,t)$.}
\label{alg:backup}
\begin{algorithmic}
\renewcommand{\algorithmicrequire}{\textbf{procedure}}
\REQUIRE Backup($x$)
\IF{$t<T$}
\STATE Compute $\mathbf{\tilde x}^0,\mathbf{\tilde u}^0$ from $x(t)=x$ using (\ref{RobustSpec})
\STATE Update $t\gets t+1$
\RETURN $\mathbf{\tilde u}^0(0)$
\renewcommand{\algorithmicrequire}{\textbf{procedure}}
\ELSE
\RETURN $\pi_{\text{stable}}(x)$
\ENDIF
\REQUIRE 
InitializeBackup($x$)
\STATE Compute target equilibrium point $z_e\gets\rho(x)$
\STATE Compute invariant set $\mathcal{G}_e$ for $z_e$
\STATE Compute $\mathbf{\bar x}^*,\mathbf{\bar u}^*$ from $x(0)=x$ using (\ref{NominalSpec2})
\STATE Initialize $t\gets 0$
\RETURN $\pi_{\text{backup}}(\cdot)=\pi_{\text{backup}}(\;\cdot\;;z_e,\mathcal{G}_e,\mathbf{\bar x}^*,\mathbf{\bar u}^*,t)$
\end{algorithmic} 
\end{algorithm}

\section{Robust Model Predictive Shielding}
\label{sec:alg}

The key challenge of implementing MPS is how to check whether a state $x$ is recoverable. When the dynamics are deterministic, this check can be performed by simulating the dynamics~\cite{DBLP:journals/corr/abs-1905-10691}. However, for stochastic dynamics, each simulation will result in different realizations of $w(k)$, so this approach no longer applies. When the dynamics are linear and the perturbations $w(k)$ are Gaussian, we can analytically check a high-probability variant of recoverability~\cite{DBLP:journals/corr/abs-1803-08552}, but this approach does not generalize to nonlinear dynamics.

Our approach is to based on robust control. First, we use tracking NMPC as the recovery policy~\cite{doi:10.1002/rnc.1758}. This choice ensures that the system reach its goal with high probability even with stochastic perturbations. Second, we check recoverability by estimating the reachable set of $\pi_{\text{backup}}$. In particular, we use the reachable set to ensure that the trajectory generated using $\pi_{\text{backup}}$ (i) is safe, and (ii) reaches an invariant set around an equilibrium point. A key innovation in our approach is that we use tools from statistical learning theory to obtain provable guarantees on our check for recoverability.

\subsection{The Backup Controller}

We use a standard robust NMPC as the backup controller~\cite{doi:10.1002/rnc.1758}. At a high level, this controller first computes a reference trajectory that transitions the system to an equilibrium point. Then, it uses NMPC to track this reference trajectory. Finally, once the trajectory has reached the invariant set $\mathcal{G}_e$ around equilibrium point $z_e\in\mathcal{X}\times\mathcal{U}$, it uses a feedback controller $\pi_{\text{stable}}$ to stabilize the system within $\mathcal{G}_e$.

\textbf{Stabilization near equilibrium points.}
We assume given a mapping $\rho:\mathcal{X}\to\mathcal{X}\times\mathcal{U}$, where $z_e=(x_e,u_e)=\rho(x)$ is an \emph{equilibrium point}---i.e., $x_e=f(x_e,u_e)$. Intuitively, $\rho(x)$ returns the equilibrium point $z_e$ ``closest'' to $x$. Then, $\pi_{\text{rec}}$ tries to transition the system from $x$ to $z_e$. Near $z_e$, we can use an LQR $\pi_{\text{stable}}$ to stabilize the system and ensure safety.

We assume we can compute a safe invariant set $\mathcal{G}_e\subseteq\mathcal{X}_{\text{safe}}$ around $z_e$---i.e., for $x\in\mathcal{G}_e$, the trajectory $\mathbf{x}$ generated using $\pi_{\text{stable}}$ from $x(0)=x$ (i) is safe, and (ii) remains in $\mathcal{G}_e$. Since the dynamics are stochastic, we cannot guarantee this property with probability $1$ (unless $w(k)$ is bounded). Nevertheless, in our experiments, we find that $\pi_{\text{stable}}$ is effective at ensuring safety and stability inside $\mathcal{G}_e$. We discuss how we compute $\mathcal{G}_e$ in Section~\ref{sec:impl}.

\textbf{Reference trajectory.}
We denote the nominal dynamics by $\bar x(k+1) = f(\bar x(k),\bar u(k))$, where $\bar x(k)$ is the nominal state and $\bar u(k)$ is the nominal control input. Given an initial state $x\in\mathcal{X}$ and a time horizon $T$, we compute a nominal trajectory to transition the system to an equilibrium point $z_e=\rho(x)$ by solving the following:
\begin{alignat}{3}
& \operatorname*{\arg\min}_{\mathbf{\bar{x}},\mathbf{\bar{u}}} ~ && \sum_{k=0}^{T-1} \ell(\bar x(k)-x_e, \bar u(k)-u_e) \label{NominalSpec} \\
& \text{subj. to} ~ && \bar x(0) = x, ~ \bar x(T) = x_e, ~ \bar u(T) = u_e, \nonumber \\
&                  && \bar x(k) \in \mathcal{\bar X}_{\text{safe}}, ~ \bar u(k) \in \mathcal{U}, \nonumber \\
&                  && \bar x(k+1) = f(\bar x(k),\bar u(k)) ~ (\forall k \in \{0,...,T-1\}) \nonumber
\end{alignat}
where $\ell(x, u) = \|x\|_Q^2 + \|u\|_R^2$ for some $Q \in \mathbb{S}^{n_X}$ and $R \in \mathbb{S}^{n_U}$. Furthermore, $\mathcal{\bar X}_{\text{safe}}\subseteq\mathcal{X}_{\text{safe}}$ can be specified by the user to improve robustness; we describe heuristics for computing these sets in Section~\ref{sec:impl}. We denote the solution to (\ref{NominalSpec}) by $(\mathbf{\bar{x}}^0,\mathbf{\bar{u}}^0)\in\mathcal{X}^{T+1}\times\mathcal{U}^{T+1}$. Since $(x_e,u_e)$ is a nominal equilibrium, the infinite horizon trajectory
\begin{align}
\label{NominalSpec2}
\mathbf{\bar x}^* = \mathbf{\bar x}^0 \circ (x_e, x_e, ...), \hspace{0.2in}
\mathbf{\bar u}^* = \mathbf{\bar u}^0 \circ (u_e, u_e, ...),
\end{align}
where $\circ$ is concatenation, is safe for the nominal dynamics.

\textbf{Tracking NMPC.}
Once we have a reference trajectory $\mathbf{\bar x}^*,\mathbf{\bar u}^*$, we use NMPC to track this reference trajectory and try to reach the equilibrium $z_e$. In particular, if we are at state $x(t)$ after $t$ steps, this controller solves the following:
\begin{alignat}{3}
\label{RobustSpec}
&\operatorname*{\arg\min}_{\mathbf{\tilde x},\mathbf{\tilde u}} ~
&& \sum_{k=0}^{T-1} \ell (\tilde x(k) - \bar x^{*}(t + k), \tilde u(k) - \bar u^{*}(t + k)) \\
&&& \hspace{0.2in} + V_f(\tilde x(T)) \nonumber \\
& \text{subj. to} ~ && \tilde x(0) = x(t), ~ \tilde x(k) \in \mathcal{X}_{\text{safe}}, ~ \tilde u(k) \in \mathcal{U}, \nonumber \\
&                   && \tilde x(k+1) = f(\tilde x(k), \tilde u(k)) ~ (\forall k \in \{0, ..., T-1\}) \nonumber
\end{alignat}
where $V_f(x)$ is the cost-to-go function of the LQR for the linearization of the nominal dynamics $f$ around $z_e$~\cite{doi:10.1002/rnc.1758}. We let $\mathbf{\tilde x}^0,\mathbf{\tilde u}^0\in\mathcal{X}^{T+1}\times\mathcal{U}^{T+1}$ be the solution of (\ref{RobustSpec}).

\textbf{Backup controller.}
Given an state $x$, an equilibrium point $z_e\in\mathcal{X}\times\mathcal{U}$, and a time horizon $T$, our backup controller $\pi_{\text{backup}}$ first computes the reference trajectory $\mathbf{\bar x}^0,\mathbf{\bar u}^0$ using (\ref{NominalSpec}), with corresponding infinite horizon reference trajectory $\mathbf{\bar x}^*,\mathbf{\bar u}^*$. Then, for each step $t\in\{0,...,T-1\}$, $\pi_{\text{backup}}$ solves (\ref{RobustSpec}) for the current state $x(t)$ to obtain $\mathbf{\tilde x}^0,\mathbf{\tilde u}^0$, and chooses control input $u(t) = \tilde{u}^0(0)$. Finally, for $t\ge T$, it chooses control input $u(t) = \pi_{\text{stable}}(x(t))$.

This procedure for computing the backup controller is summarized in Algorithm~\ref{alg:backup}. Note that $\pi_{\text{backup}}$ actually needs to keep internal state consisting of the target equilibrium point $z_e=\rho(x)$, its corresponding invariant set $\mathcal{G}_e$, the reference trajectory $\mathbf{\bar x}^*,\mathbf{\bar u}^*$ to the equilibrium point, and the number of steps $t$ taken so far using the backup controller. This internal state is initialized in the context of a given state $x\in\mathcal{X}$ by the function call InitializeBackup($x$).

\subsection{Checking Robust Recoverability via Sampling}
\label{sec:montecarlo}

Building on ideas from tube NMPC~\cite{book}, we use sampling to determine whether $\pi_{\text{rec}}$ can safely reach the invariant set $\mathcal{G}_e$ from the current state $x$. In particular, we sample $N$ trajectories according to the stochastic dynamics, and fit boxes $B(t)$ that cover all the states sampled on each given step $t\in\{0,...,T\}$. Intuitively, if we take $N$ to be sufficiently large, then the realized trajectory will lie in $B(t)$ at each step $t$ with high probability. In contrast to prior work, we make this intuition precise using tools from statistical learning theory. Finally, to check if $x$ is recoverable, we check that it is robustly safe according to the uncertainty in these boxes, and furthermore that it robustly enters the invariant set $\mathcal{G}_e$.

\begin{algorithm}[t]
\caption{Check if $x$ is robustly recoverable.}
\label{alg:recoverable}
\begin{algorithmic}
\renewcommand{\algorithmicrequire}{\textbf{procedure}}
\REQUIRE 
IsRecoverable($x$)
\STATE $\pi_{\text{backup}}\gets\text{InitializeBackup}(x)$
\STATE Let $\mathbf{\bar x}^*$ be the reference trajectory of $\pi_{\text{backup}}$
\STATE $\mathbf{B}\gets\text{EstimateReachableSets}(x,\pi_{\text{backup}})$
\FOR{$t\in\{0,...,T\}$}
\IF{$\bar x^*(t)\not\in\mathcal{X}_{\text{safe}}\ominus B(t)$}
\RETURN false
\ENDIF
\ENDFOR
\IF{$\bar x^*(T)\not\in\mathcal{G}_e\ominus B(T)$}
\RETURN false
\ENDIF
\RETURN true
\renewcommand{\algorithmicensure}{\textbf{Output:}}
\end{algorithmic} 
\end{algorithm}

\begin{algorithm}[t]
\caption{Estimates the reachable sets $B(t)$ after $t$ steps using Monte Carlo sampling.}
\begin{algorithmic}
\label{alg:getTightenedSet}
\renewcommand{\algorithmicrequire}{\textbf{procedure}}
\REQUIRE EstimateReachableSets($x, \pi_{\text{backup}}$)
\STATE Compute $N$ that satisfies (\ref{sampleNumber})
\FOR{$i\in[N]$}
\STATE Sample $w\sim\mathcal{P}_{\mathcal{W}}$, and let $x'=x+w$
\STATE Sample $(x_i(0),...,x_i(T))$ from $x_i(0)=x'$ using $\pi_{\text{backup}}$
\ENDFOR
\FOR{$t\in\{0,1,...,T\}$}
\STATE Fit $B(t)\in\mathcal{B}$ to $\{x_i(t)\mid i\in[N]\}$
\ENDFOR
\RETURN $\mathbf{B}=(B(0),...,B(T))$
\renewcommand{\algorithmicensure}{\textbf{Output:}} 
\end{algorithmic} 
\end{algorithm}

\textbf{Robust recoverability.}
Our goal is to ensure that $\pi_{\text{rec}}$ can always transition the system safely from the current state $x$ to the invariant set $\mathcal{G}_e$ around $z_e=\rho(x)$. Due to the random perturbation $w(k)$ in the dynamics, we cannot make an absolute guarantee that this property holds. Following prior work~\cite{gillula2012guaranteed,akametalu2014reachability,berkenkamp2017safe}, we instead aim to guarantee that this property holds with high probability.
\begin{definition}
\rm
Let $\epsilon\in\mathbb{R}_{>0}$ and $x\in\mathcal{X}$ be given. Let $z_e=\rho(x)$, let $\mathcal{G}_e$ be an invariant set for $\pi_{\text{stable}}$ around $z_e$, and let $\mathbf{x}^0=(x(0),...,x(T))$ be a trajectory generated using $\pi_{\text{backup}}$ from $x(0)=x+w$, where $w\sim\mathcal{P}_{\mathcal{W}}$. We say $x$ is $\epsilon$ \emph{robustly recoverable} if with probability at least $1-\epsilon$, (i) $x(t)$ is safe for every $t\in\{0,...,T\}$, and (ii) $x(T)\in\mathcal{G}_e$.
\end{definition}

In other words, $\mathbf{x}^0$ safely transitions the system from $x$ to $\mathcal{G}_e$ with probability at least $1-\epsilon$. Algorithm~\ref{alg:recoverable} checks whether a state $x$ is robustly recoverable using sampling. Prior work on safe reinforcement learning~\cite{gillula2012guaranteed,akametalu2014reachability,berkenkamp2017safe} has relied on thresholding the perturbation distribution and then using verification to obtain similar bounds. This approach can guarantee that $x$ is robustly recoverable with probability $1$. In contrast, using our approach, there is an additional chance $\delta$ (for any given $\delta\in\mathbb{R}_{>0}$) that our algorithm incorrectly concludes that $x$ is robustly recoverable when it is not. The difference is that the $\epsilon$ error in robust recoverability is due to the noise in the dynamics, whereas our $\delta$ error is due to noise in the sampled trajectories taken by our algorithm.

We believe this added potential for error is reasonable for two reasons. First, there is already an $\epsilon$ chance of error, so the added error $\delta$ does not affect the nature of our guarantee. Second, the dependence of the running time of our algorithm $1/\delta$ is logarithmic, so we can choose $\delta$ to be very small.

\textbf{Estimating reachable sets.}
Our approach is to compute sets $B(t)$ for $t\in\{0,...,T\}$ such that the trajectory $\textbf{x}^0$ satisfies $x(t)-\bar x^*(t)\in B(t)$ with probability at least $1-\epsilon$, where $\mathbf{\bar x}^*$ is the reference trajectory used by $\pi_{\text{backup}}$---i.e.,
\begin{align}
\label{eqn:robustcontains}
\text{Pr}(x(t)-\bar x^*(t)\in B(t))\ge1-\epsilon,
\end{align}
where the probability is taken over the randomness in the dynamics. To this end, a \emph{box constraint} is a set
\begin{align*}
B=[a_1,b_1]\times...\times[a_n,b_n]\subseteq\mathbb{R}^n,
\end{align*}
where $[a,b]\subseteq\mathbb{R}$ denotes the closed interval from $a$ to $b$. We use $\mathcal{B}$ to denote the set of all possible boxes. Now, we have the following theoretical guarantee.
\begin{lemma}
\label{lem:probGuarantee}
Let $d$ be a distribution over $\mathbb{R}^n$ and $\epsilon,\delta\in\mathbb{R}_{>0}$ be given. Consider $N$ i.i.d. samples $x_1,...,x_N\sim d$, where
\begin{align}
\label{sampleNumber}
N \geq \frac{n\log{\frac{2eN}{n}} + \log{\frac{4}{\delta}}}{\epsilon^2},
\end{align}
and let $B\in\mathcal{B}$ be any box satisfying $x_i\in B$ for all $i\in\{1,...,N\}$. Then, with probability at least $1-\delta$, we have
\begin{align}
\label{PAC}
\text{Pr}_{x\sim d}(x\in B)\ge1-\epsilon.
\end{align}
\end{lemma}

Intuitively, (\ref{PAC}) says that at least a $1-\epsilon$ fraction of states (weighted by $d$) fall inside $B$, and this guarantee holds with probability at least $1-\delta$. The proof, based on tools from statistical learning theory, is given in Appendix A.

Algorithm~\ref{alg:getTightenedSet} takes $N$ samples of the trajectory $\mathbf{x}^0$ by simulating the dynamics, and fits a box $B(t)$ based on the sampled states on each step $t\in\{0,...,T\}$. The following guarantee follows from Lemma~\ref{lem:probGuarantee} via a union bound:
\begin{lemma}
\label{lem:estimate}
Let $\mathbf{B}$ be the sequence of boxes returned by Algorithm~\ref{alg:getTightenedSet}. With probability at least $1-(T+1)\delta$, we have
\begin{align*}
\text{Pr}(\forall t\in\{0,...,T\},~x(t)-\bar x^*(t)\in B(t))\ge1-(T+1)\epsilon.
\end{align*}
\end{lemma}
\vspace{5pt}

As before, the $1-(T+1)\delta$ probability is according to the samples used by our algorithm, whereas the $1-(T+1)\epsilon$ probability is according to the randomness in the dynamics.

\begin{figure}[t]
\centering
\includegraphics[scale=0.45]{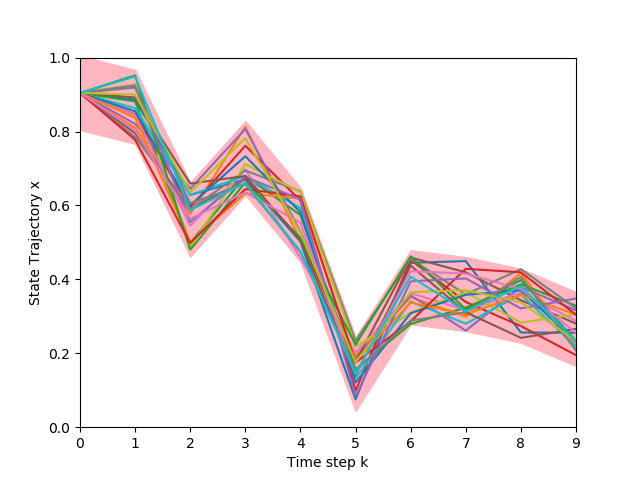}
\caption{An example of a tube (red region) estimated using Algorithm~\ref{alg:getTightenedSet} for $\pi_{\text{backup}}$. We guarantee the realized trajectory lies in this tube with high probability.}
\label{fig:Tube}
\end{figure}

The sets $B(t)$ computed using Algorithm~\ref{alg:getTightenedSet} can be thought of as an estimate of a tube in which the trajectories are guaranteed to stay~\cite{doi:10.1002/rnc.1758}. In contrast to prior work, we have used results from statistical learning theory to obtain probabilistic guarantees on the correctness of these tubes~\cite{vapnik2013nature}. An example of an estimated tube is shown in Fig.~\ref{fig:Tube}. 

\textbf{Checking recoverability.}
Given the boxes $B(t)$ for $t\in\{0,...,T\}$, Algorithm~\ref{alg:recoverable} checks both properties required for robust recovery: (i) to check if $x(t)\in\mathcal{X}_{\text{safe}}$ with high probability, it checks if
\begin{align*}
\{\bar x^*(t)+x\mid x\in B(t)\}\subseteq\mathcal{X}_{\text{safe}},
\end{align*}
which is equivalent to $\bar x^*(t)\in\mathcal{X}_{\text{safe}}\ominus B(t)$, and (ii) to check if $x(T)\in\mathcal{G}_e$ with high probability, it checks if
\begin{align*}
\{\bar x^*(t)+x\mid x\in B(T)\}\subseteq\mathcal{G}_e,
\end{align*}
which is equivalent to $x(T)\in\mathcal{G}_e\ominus B(T)$. These checks ensure robust recoverability because Corollary~\ref{lem:estimate} ensures that $x(t)\in B(t)$ with high probability for every $t\in\{0,...,T\}$. Thus, we have the following guarantee:
\begin{lemma}
\label{lem:recoverable}
Given a state $x\in\mathcal{X}$, if Algorithm~\ref{alg:recoverable} returns true, then $x$ is $(T+1)\epsilon$ robustly recoverable with probability $1-(T+1)\delta$ (according to the randomness in the algorithm).
\end{lemma}

\subsection{Robust Model Predictive Shielding}
\label{sec:rmpsdetails}

Our robust model predictive shielding (RMPS) algorithm is shown in Algorithm~\ref{alg:shield}.
At state $x\in\mathcal{X}$, this algorithm computes a control input (denoted $\pi_{\text{shield}}(x)$) by checking whether next state $f^{(\hat{\pi})}(x)$ is robustly recoverable (with high probability) in simulation. Otherwise, it takes a step according to $\pi_{\text{backup}}$. One subtlety is that if $\pi_{\text{backup}}$ has already been initialized, it actually needs to check if $f^{(\pi_{\text{backup}})}(x)$ is robustly recoverable. The issue is that robust recoverability is defined with respect to a freshly initialized backup policy, \emph{not} the backup policy after it has taken some number of steps. We have the following guarantee:
\begin{theorem}
\label{thm:main}
If $x\in\mathcal{X}$ is $1-(T+1)\epsilon$ robustly recoverable, then $f^{(\pi_{\text{shield}})}(x)+w$ (where $w\sim\mathcal{P}_{\mathcal{W}}$) is $1-(T+1)\epsilon$ robustly recoverable with probability at least $1-2(T+1)\delta$.
\end{theorem}

See Appendix~\ref{sec:mainproof} for a proof. A key shortcoming of this guarantee is that it does not ensure safety of the infinite horizon trajectory. Given our assumptions, a stronger guarantee is impossible, since on every step there is a chance that the additive perturbation is large, causing the system to leave $\mathcal{X}_{\text{safe}}$. However, this guarantee is still useful since it helps guide the design of our algorithm. In practice, we find that the bounds can be tighter than the theory suggests, since the robust NMPC is actually conservatively overapproximating the reachable set. In other words, the robust NMPC ensures safety much more robustly than the probabilities in Theorem~\ref{thm:main} would suggest.

\begin{figure*}[t]
    \centering
    \includegraphics[scale=0.18]{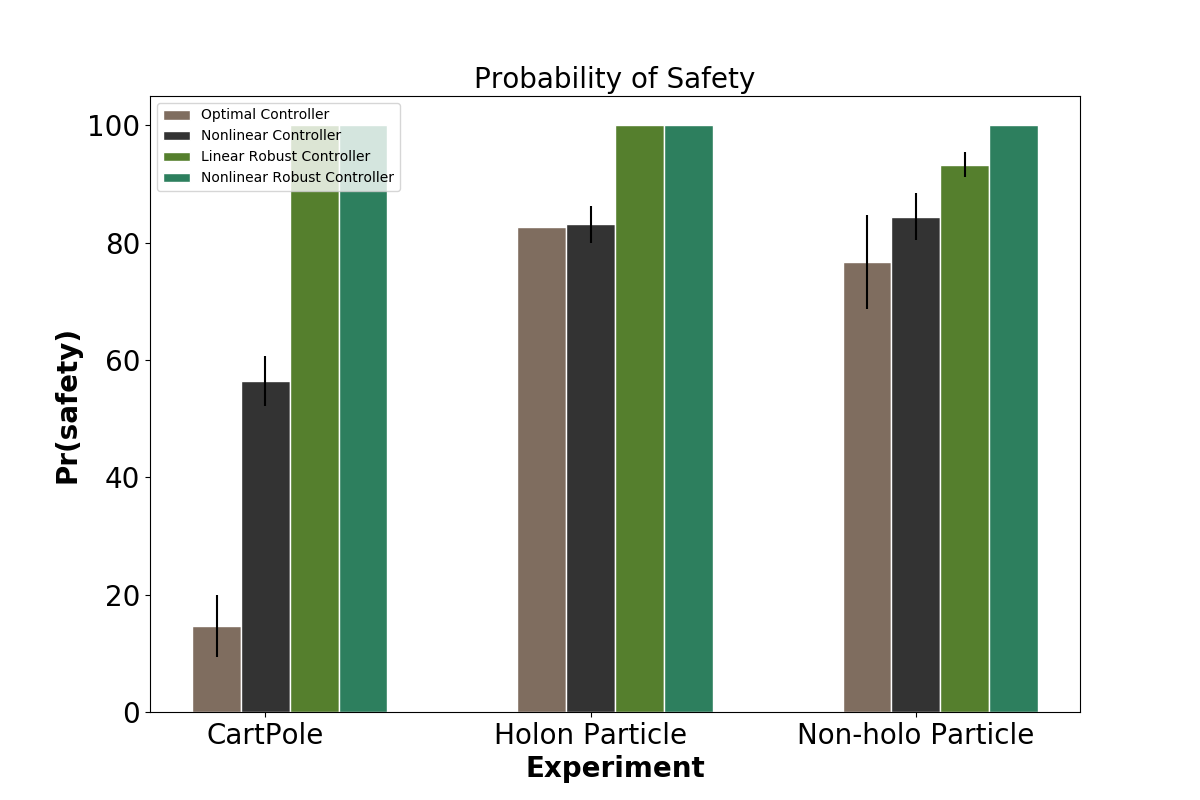}
    \includegraphics[scale=0.18]{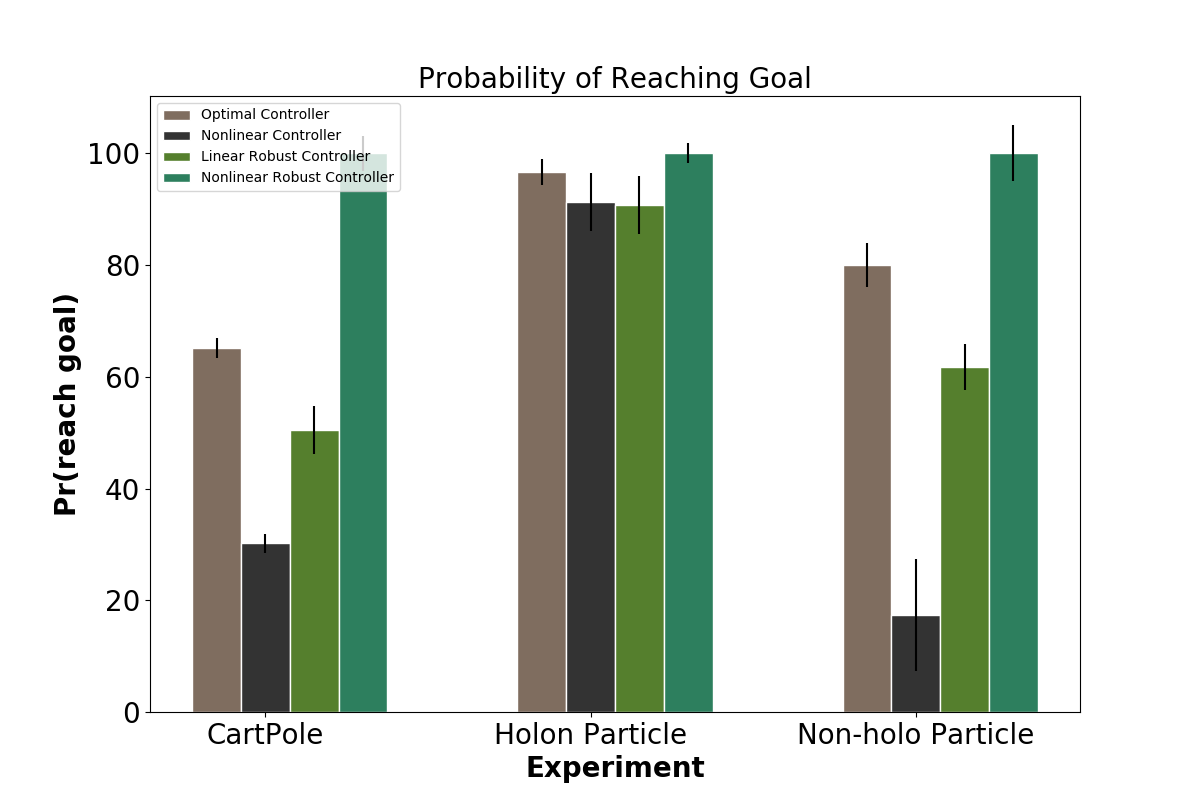}
    \includegraphics[scale=0.18]{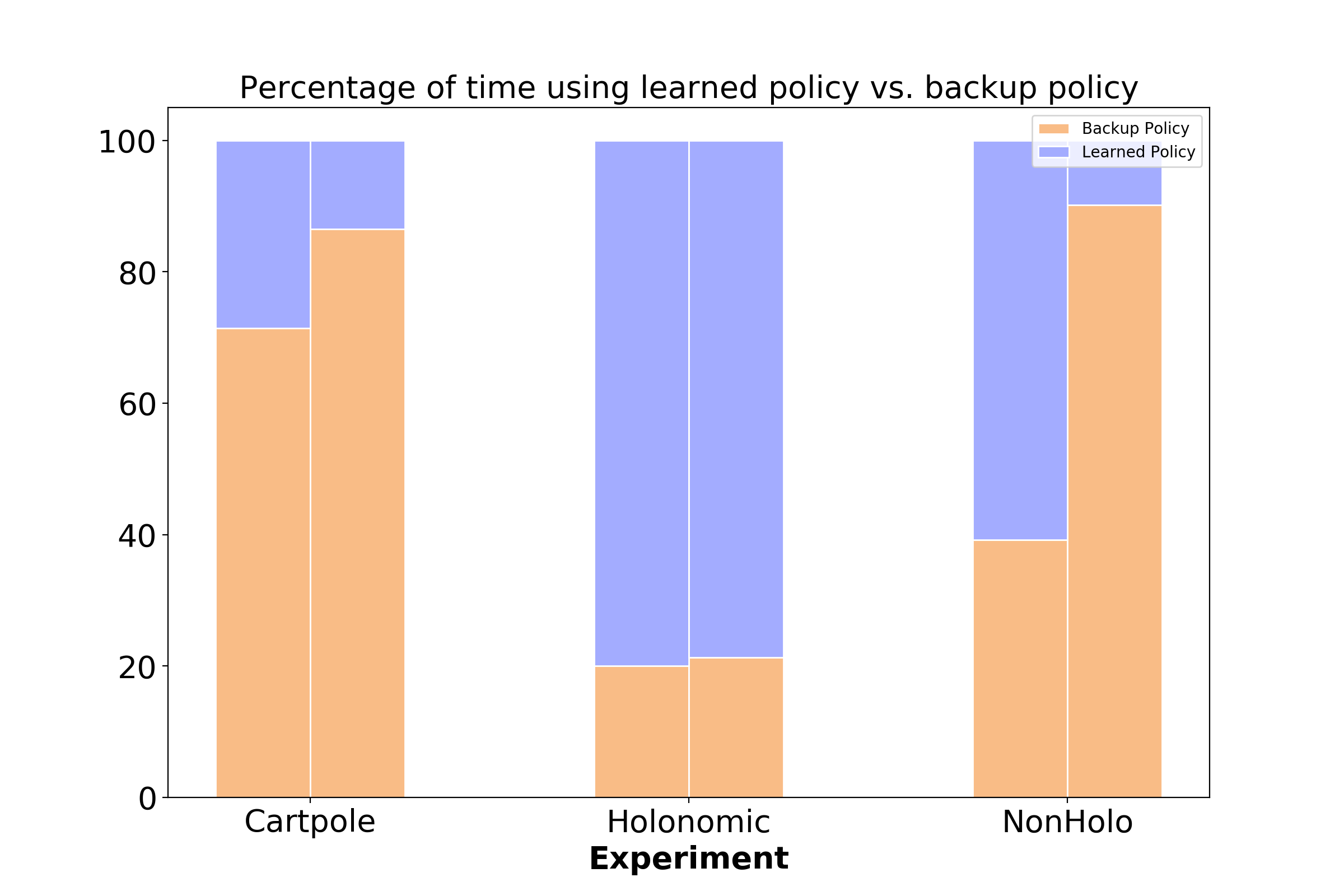}
    \caption{Left: Safety probabilities for no shielding (light brown), non-robust MPS (dark brown), LRMPS (light green), and RMPS (dark green). Middle: Probability of reaching goal for the same policies. Right: Percentage of time using the learned policy $\hat\pi$ (blue) compared to the backup policy $\pi_{\text{backup}}$ (orange) by RMPS (left) and LRMPS (right).}
    \label{fig:prob}
\end{figure*}

\subsection{Running Time}

We briefly discuss the running time of Algorithm~\ref{alg:shield}. First, it is straightforward to see that the number of samples $N$ is $\tilde{O}(1/\epsilon^2)$ (where $\tilde{O}$ indicates that we have suppressed log factors including $\log(1/\delta)$). For each sample, we need to simulate the closed-loop dynamics $f^{(\pi_{\text{rec}})}$ over a horizon of length $T$. Thus, assuming the cost of simulating $f^{(\pi_{\text{rec}})}$ for one step is $\tau$, the running time of Algorithm~\ref{alg:shield} is $\tilde{O}(T\tau/\epsilon^2)$.

\subsection{Practical Modifications}
\label{sec:impl}

We describe several practical modifications to our algorithm designed to improve either performance or computational tractability. These modifications may weaken our safety guarantees, but as we show in our experiments, they do not affect safety very much empirically.

\textbf{Computing $\mathcal{G}_e$.}
We use a heuristic to compute $\mathcal{G}_e$ from~\cite{doi:10.1002/rnc.1758}. In particular, we sample trajectories over a long horizon and estimate the reachable set $B$ the same way as in Algorithm~\ref{alg:getTightenedSet}. This approach does not provide any guarantees that the estimated set $\mathcal{G}_e$ is actually invariant, but it works well in our experiments.

\textbf{Precomputing $B$.}
Computing the sets $B(t)$ (for $t\in\{0,...,T\}$) on-the-fly can be prohibitively expensive, since we might need a large number of samples $N$ for Lemma~\ref{lem:probGuarantee} to apply. Instead, we precompute these sets from a fixed initial state $x_0$. Then, we reuse the same states rather than recomputing them at each step. Intuitively, this approach works well in practice since the dynamics of the tracking NMPC are usually fairly similar for different initial states.


\textbf{Using tighter constraints for NMPC.}
In the optimization problem (\ref{NominalSpec}) used to compute the reference trajectory, we noted that we can use tighter state and input constraints $\bar{x}(k)\in\bar{\mathcal{X}}_{\text{safe}}$ than needed. In particular, by doing so, we can improve the robustness of the tracking NMPC---we use the ``tightened set'' $\bar{\mathcal{X}}_{\text{safe}}=\mathcal{X}_{\text{safe}}\ominus B(t)$. Unlike the previous two heuristics, this one does not affect our theoretical guarantees, since our guarantees hold for an arbitrary backup policy. While our approach would ensure safety even without the tighter state and input constraints, using tighter constraints improves the chances that the backup policy recovers the system from a given state $x$---i.e., it improves the size of $\mathcal{X}_{\text{rec}}$. In other words, it increases the probability that the robot reaches its goal without diminishing the probability of safety.

\section{Experiments}
\label{sec:Exp}

We perform experiments to demonstrate how our system can ensure safety of stochastic systems with nonlinear dynamics and/or nonconvex constraints.

\subsection{Setting}

We perform experiments using three environments: (i) cart-pole, which has nonlinear dynamics and polytopic constraints (i.e., the pole should not fall below a certain height), a particle with holonomic dynamics and obstacles (which has linear dynamics but nonconvex constraints), and a particle with non-holonomic dynamics and obstacles (which has both nonlinear dynamics and nonconvex constraints).

For the cart-pole, the states are $z=(x,v,\theta,\omega)\in\mathbb{R}^4$, where $x$ is the cart position, $v$ is the cart velocity, $\theta$ is the pole angle from upright position, and $\omega$ is the pole angular velocity, and the control inputs are $u\in\mathbb{R}$, with the goal of reaching a target position $x_{\text{target}}$~\cite{DBLP:journals/corr/BrockmanCPSSTZ16}. The safety constraint is that the pole should not fall down while moving the cart. We define the cost function to be
\begin{align*}
\ell(z,u) = (x - x_{\text{target}})^2 + \gamma\cdot\theta^2,  
\end{align*}
where $\gamma\in\mathbb{R}_{>0}$ is a hyperparameter. Finally, disturbances are uniform noise $w_i(k)\sim\text{Uniform}([-0.01, 0.01])$ for the velocity and angular velocity, and zero otherwise.

For the single particle with holonomic dynamics, the states are $z = (x, y, v_x, v_y)\in\mathbb{R}^4$, where $(x,y)$ is position and $(v_x, v_y)$ is velocity, and the control inputs are $(u_x, u_y)\in\mathbb{R}^2$, where $(u_x,u_y)$ is the acceleration. The system dynamics are $z(k+1)=Az(k)+Bu(k)$, where
\begin{align*}
A= \begin{bmatrix} 
0 & 0 & 1 & 0\\
0 & 0 & 0 & 1\\
0 & 0 & 0 & 0\\
0 & 0 & 0 & 0
\end{bmatrix},
B = \begin{bmatrix} 
0 & 0 \\
0 & 0 \\
1 & 0 \\
0 & 1
\end{bmatrix}.
\end{align*}
The cost function is
\begin{align*}
 \ell(z,u) = -d((x,y),g)^2 + \gamma\cdot \sum_{i=1}^N d((x,y),o)^2,
\end{align*}
where $g\in\mathbb{R}^2$ is the goal, $o_i$ ($i\in[N]$) are the obstacles, and $d(x,y)=\|x-y\|_2$, and $\gamma$ is a hyperparameter.  Disturbances are uniform noise $w_i(k)\sim\text{Uniform}([-0.01, 0.01])$ for the velocity and angular velocity, and zero otherwise.

For the single particle with non-holonomic dynamics, the states are $z = (x, y, v, h)\in\mathbb{R}^4$, where $(x,y)$ is the position, $v$ is the velocity, and $h$ is the heading, and the control inputs are $(a, \omega)\in\mathbb{R}^2$, where $a$ is acceleration and $\omega$ is angular acceleration. The system dynamics are
\begin{align*}
f(z,u)=z+(v\cdot\cos(h),v\cdot\sin(h),a,v\cdot\tan(\omega)/\ell)
\end{align*}
where $\ell$ is the particle radius. Costs and disturbances are the same as for the holonomic particle.

We compare RMPS with: (i) using the learned policy $\hat{\pi}$ without any shielding, (ii) using shielding without robust control~\cite{DBLP:journals/corr/abs-1905-10691}, which we call non-robust MPS, and (iii) using RMPS with a linear robust controller~\cite{DBLP:journals/corr/BrockmanCPSSTZ16}, which we call LRMPS. In particular, for LRMPS, we linearize the dynamics around the origin, and then use the corresponding linear robust MPC proposed in~\cite{DBLP:journals/corr/BrockmanCPSSTZ16} to try and recover the system. For each experiment, we run 50 scenarios with 3 different random seeds and compute both the probability of safety and the probability of reaching the goal. Also, we use $N=1500$ to estimate the tightened constraints in NMPC.

\subsection{Results}

The safety and performance of the four algorithms are shown in Fig.~\ref{fig:prob}. First, as can be seen, RMPS and LRMPS achieve a safety probability of 1.0. In contrast, the learned policy is frequently unsafe, demonstrating the importance of shielding. Furthermore, RMPS only slightly diminishes performance compared to the learned policy. Similarly, the non-robust MPS performs slightly better than the learned policy alone in terms of safety, but still cannot guarantee that safety holds. In contrast, RMPS guarantees safety in each environment. Its performance is slightly diminished---for the particle with non-holonomic dynamics (the hardest environment), the probability of reaching the goal drops by about 20\%. Thus, RMPS is much more suitable for safety-critical systems where safety must be guaranteed.

\begin{figure}
    \centering
    \includegraphics[scale=0.4]{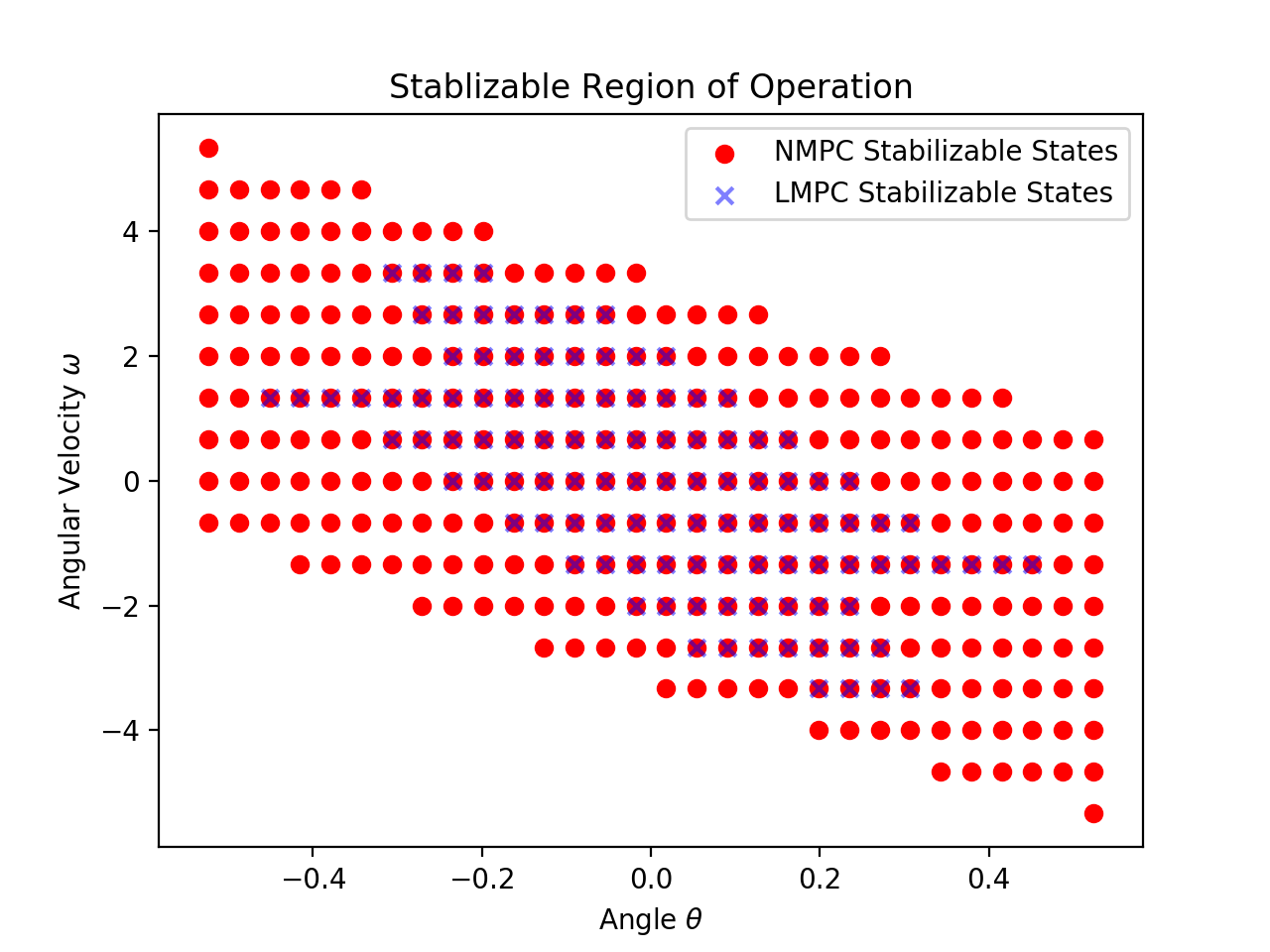}
    \caption{Pairs $(\theta, \omega)$ for which nonlinear MPC (red dot) and linear MPC (blue cross) can steer the pole to upright position.}
    \label{stable region}
\end{figure}

Next, except for the holonomic particle (which has linear dynamics), RMPS achieves much better performance than the LRMPS. This improvement in performance comes from the fact that our NMPC can stabilize the system in many more states compared to the linear robust MPC. For example, as shown in Fig.~\ref{stable region}, for a fixed cart speed, NMPC can recover the pole to the upright position for more $(\theta, \omega)$ pairs compared to linear MPC. Thus, RMPS is less conservative than LRMPS and has better performance. Fig.~\ref{fig:prob} (right) also supports these results---LRMPS uses backup controller more frequently than RMPS, showing that it is more conservative. These results demonstrate the benefit of accounting for the nonlinearity in the dynamics.

\section{Conclusion}

We have proposed an algorithm for ensuring the safety of a learned policy for a stochastic nonlinear dynamical system. We use a sampling-based approach to estimate the reachable set of the backup policy, and use results from statistical learning theory to provide theoretical guarantees on our estimates. In our experiments, we show that our approach can ensure safety without sacrificing very much performance despite these modifications. Thus, our approach is a promising way to ensure safety in safety-critical systems.

A key direction for future work is reducing the running time of our algorithm. While our approach is already computationally tractable, it is still too slow for production systems. As we described, our current implementation uses two heuristics to improve performance, but these heuristics break our theoretical guarantees. Thus, improving the performance of our algorithm would enable our algorithm to be used with its theoretical guarantees intact. Additional directions for future work include enabling safe exploration and extending our results to partially observable systems, as well as demonstrating our approach on real robotics systems.

\section*{Appendix} \label{APPENDIX}

\subsection{Proof of Lemma~\ref{lem:probGuarantee}} \label{prf:th2}

Given $B\in\mathcal{B}$, define $f_B:\X\to\mathcal{Y}$, where $\mathcal{Y}=\{0,1\}$, by $f_B(x)=\mathbb{I}(x\in B)$---i.e., $f_B$ indicates whether $x$ is contained in $B$. Note that $f_B$ is a binary classifier, and we can consider the family of classifiers $\F=\{f_B\mid B\in\mathcal{B}\}$. also, define the distribution $\tilde{d}$ on $\mathcal{X}\times\mathcal{Y}$ by $\tilde{d}(x,y)=d(x)\cdot\mathbb{I}(y=1)$---i.e., all labels are $1$. Thus, sampling $x\sim d$ is equivalent to sampling $(x,y)\sim\tilde{d}$, where $y=1$. Recall that we choose $B$ so all samples $x_1,...,x_N\sim d$ satisfy $x_i\in B$. Equivalently, $(x_1,y_1),...,(x_N,y_N)\sim\tilde{d}$, where $y_i=1$ for all $i\in[N]$, so $f_B(x_i)=y_i$ for all $i\in[N]$. Thus, we can think of choosing $B$ as choosing $f_B\in\mathcal{F}$ such that the training error 
\begin{align*}
\hat{L}(f_B)=N^{-1}\sum_{i=1}^N\mathbb{I}(f_B(x_i)\neq y_i)=0
\end{align*}
on a set of i.i.d. samples $(x_i,y_i)\sim d$. Thus, we can apply results from statistical learning theory to bound the test error
\begin{align*}
L^*(f_B)=\mathbb{E}_{(x,y)\sim\tilde{d}}(f_B(x)=y)=1-\text{Pr}_{x\sim d}(x\in B),
\end{align*}
where the last probability is the one we are seeking to bound. In particular, it is straightforward to check that the VC dimension of $f_B$ for boxes $B\subseteq\mathbb{R}^n$ is $\text{VC}(\mathcal{F})=n$. By the VC dimension bound~\cite{vapnik2013nature}, for all $B\in\mathcal{B}$, we have
\begin{align}
\label{eqn:vcdim}
L^*(f_B)
& \leq \sqrt{\frac{1}{N}\left(n\log \frac{2eN}{n} + \log \frac{4}{\delta}\right)}
\end{align}
with probability at least $1-\delta$. The claim follows by setting $\epsilon$ equal to the right-hand side of (\ref{eqn:vcdim}). $\blacksquare$

\subsection{Proof of Theorem~\ref{thm:main}} \label{sec:mainproof}

If $\pi_{\text{shield}}$ uses $\hat{\pi}$, by Lemma~\ref{lem:recoverable}, $f^{(\hat{\pi})}(x)$ is $1-(T+1)\epsilon$ robustly recoverable with probability $\ge1-(T+1)\delta$. If $\pi_{\text{shield}}$ uses the already initialized version of $\pi_{\text{backup}}$, by Lemma~\ref{lem:recoverable}, $f^{(\pi_{\text{backup}})}(x)$ is $1-(T+1)\epsilon$ robustly recoverable with probability $\ge1-(T+1)\delta$. If $\pi_{\text{shield}}$ initializes $\pi_{\text{backup}}$ and uses it, since $x$ is $1-(T+1)\epsilon$ robustly recoverable,
\begin{align*}
\phi_0=(\forall t\in\{0,1,...,T\},~x(t)\in\mathcal{X}_{\text{safe}}\wedge x(T)\in\mathcal{G}_e)
\end{align*}holds with probability at least $1-(T+1)\epsilon$. The robust recoverability condition for $x(1)=f^{(\pi_{\text{backup}})}(x)$ is
\begin{align*}
\phi_1=(\forall t\in\{1,...,T\},~x(t)\in\mathcal{X}_{\text{safe}}\wedge x(T)\in\mathcal{G}_e).
\end{align*}
In particular, note that $\phi_0\Rightarrow\phi_1$, so
\begin{align*}
\text{Pr}(\phi_1)\ge\text{Pr}(\phi_0)\ge1-(T+1)\epsilon,
\end{align*}
so $f^{(\pi_{\text{shield}})}(x)$ is $1-(T+1)\epsilon$ robustly recoverable. By a union bound, $f^{(\pi_{\text{shield}})}(x)$ is robustly recoverable with probability $\ge1-2(T+1)\delta$, as claimed. $\blacksquare$

\bibliographystyle{IEEEtran}
\bibliography{IEEEexample}

\begin{thebibliography}{10}
\providecommand{\url}[1]{#1}
\csname url@rmstyle\endcsname
\providecommand{\newblock}{\relax}
\providecommand{\bibinfo}[2]{#2}
\providecommand\BIBentrySTDinterwordspacing{\spaceskip=0pt\relax}
\providecommand\BIBentryALTinterwordstretchfactor{4}
\providecommand\BIBentryALTinterwordspacing{\spaceskip=\fontdimen2\font plus
\BIBentryALTinterwordstretchfactor\fontdimen3\font minus
  \fontdimen4\font\relax}
\providecommand\BIBforeignlanguage[2]{{%
\expandafter\ifx\csname l@#1\endcsname\relax
\typeout{** WARNING: IEEEtran.bst: No hyphenation pattern has been}%
\typeout{** loaded for the language `#1'. Using the pattern for}%
\typeout{** the default language instead.}%
\else
\language=\csname l@#1\endcsname
\fi
#2}}

\bibitem{DBLP:journals/corr/HasseltGS15}
\BIBentryALTinterwordspacing
H.~van Hasselt, A.~Guez, and D.~Silver, ``Deep reinforcement learning with
  double q-learning,'' \emph{CoRR}, vol. abs/1509.06461, 2015. [Online].
  Available: \url{http://arxiv.org/abs/1509.06461}
\BIBentrySTDinterwordspacing

\bibitem{DBLP:journals/corr/SchulmanLMJA15}
\BIBentryALTinterwordspacing
J.~Schulman, S.~Levine, P.~Moritz, M.~I. Jordan, and P.~Abbeel, ``Trust region
  policy optimization,'' \emph{CoRR}, vol. abs/1502.05477, 2015. [Online].
  Available: \url{http://arxiv.org/abs/1502.05477}
\BIBentrySTDinterwordspacing

\bibitem{DBLP:journals/corr/SchulmanWDRK17}
\BIBentryALTinterwordspacing
J.~Schulman, F.~Wolski, P.~Dhariwal, A.~Radford, and O.~Klimov, ``Proximal
  policy optimization algorithms,'' \emph{CoRR}, vol. abs/1707.06347, 2017.
  [Online]. Available: \url{http://arxiv.org/abs/1707.06347}
\BIBentrySTDinterwordspacing

\bibitem{DBLP:journals/corr/MnihBMGLHSK16}
\BIBentryALTinterwordspacing
V.~Mnih, A.~P. Badia, M.~Mirza, A.~Graves, T.~P. Lillicrap, T.~Harley,
  D.~Silver, and K.~Kavukcuoglu, ``Asynchronous methods for deep reinforcement
  learning,'' \emph{CoRR}, vol. abs/1602.01783, 2016. [Online]. Available:
  \url{http://arxiv.org/abs/1602.01783}
\BIBentrySTDinterwordspacing

\bibitem{44806}
\BIBentryALTinterwordspacing
D.~Silver, A.~Huang, C.~J. Maddison, A.~Guez, L.~Sifre, G.~van~den Driessche,
  J.~Schrittwieser, I.~Antonoglou, V.~Panneershelvam, M.~Lanctot, S.~Dieleman,
  D.~Grewe, J.~Nham, N.~Kalchbrenner, I.~Sutskever, T.~Lillicrap, M.~Leach,
  K.~Kavukcuoglu, T.~Graepel, and D.~Hassabis, ``Mastering the game of go with
  deep neural networks and tree search,'' \emph{Nature}, vol. 529, pp.
  484--503, 2016. [Online]. Available:
  \url{http://www.nature.com/nature/journal/v529/n7587/full/nature16961.html}
\BIBentrySTDinterwordspacing

\bibitem{DBLP:journals/corr/abs-1811-12927}
\BIBentryALTinterwordspacing
R.~Mahjourian, N.~Jaitly, N.~Lazic, S.~Levine, and R.~Miikkulainen,
  ``Hierarchical policy design for sample-efficient learning of robot table
  tennis through self-play,'' \emph{CoRR}, vol. abs/1811.12927, 2018. [Online].
  Available: \url{http://arxiv.org/abs/1811.12927}
\BIBentrySTDinterwordspacing

\bibitem{DBLP:journals/corr/abs-1901-03737}
\BIBentryALTinterwordspacing
N.~O. Lambert, D.~S. Drew, J.~Yaconelli, R.~Calandra, S.~Levine, and K.~S.~J.
  Pister, ``Low level control of a quadrotor with deep model-based
  reinforcement learning,'' \emph{CoRR}, vol. abs/1901.03737, 2019. [Online].
  Available: \url{http://arxiv.org/abs/1901.03737}
\BIBentrySTDinterwordspacing

\bibitem{DBLP:journals/corr/BojarskiTDFFGJM16}
\BIBentryALTinterwordspacing
M.~Bojarski, D.~D. Testa, D.~Dworakowski, B.~Firner, B.~Flepp, P.~Goyal, L.~D.
  Jackel, M.~Monfort, U.~Muller, J.~Zhang, X.~Zhang, J.~Zhao, and K.~Zieba,
  ``End to end learning for self-driving cars,'' \emph{CoRR}, vol.
  abs/1604.07316, 2016. [Online]. Available:
  \url{http://arxiv.org/abs/1604.07316}
\BIBentrySTDinterwordspacing

\bibitem{andrychowicz2018learning}
M.~Andrychowicz, B.~Baker, M.~Chociej, R.~Jozefowicz, B.~McGrew, J.~Pachocki,
  A.~Petron, M.~Plappert, G.~Powell, A.~Ray, \emph{et~al.}, ``Learning
  dexterous in-hand manipulation,'' \emph{arXiv preprint arXiv:1808.00177},
  2018.

\bibitem{DBLP:journals/corr/abs-1907-05300}
\BIBentryALTinterwordspacing
A.~Khan, C.~Zhang, S.~Li, J.~Wu, B.~Schlotfeldt, S.~Y. Tang, A.~Ribeiro,
  O.~Bastani, and V.~Kumar, ``Learning safe unlabeled multi-robot planning with
  motion constraints,'' \emph{CoRR}, vol. abs/1907.05300, 2019. [Online].
  Available: \url{http://arxiv.org/abs/1907.05300}
\BIBentrySTDinterwordspacing

\bibitem{khan2019graph}
A.~Khan, E.~Tolstaya, A.~Ribeiro, and V.~Kumar, ``Graph policy gradients for
  large scale robot control,'' in \emph{CoRL}, 2019.

\bibitem{tolstaya2019learning}
E.~Tolstaya, F.~Gama, J.~Paulos, G.~Pappas, V.~Kumar, and A.~Ribeiro,
  ``Learning decentralized controllers for robot swarms with graph neural
  networks,'' in \emph{CoRL}, 2019.

\bibitem{article}
C.~Gehring, S.~Coros, M.~Hutler, D.~Bellicoso, H.~Heijnen, R.~Diethelm,
  M.~Bloesch, P.~Fankhauser, J.~Hwangbo, M.~Hoepflinger, and R.~Siegwart,
  ``Practice makes perfect: An optimization-based approach to controlling agile
  motions for a quadruped robot,'' \emph{IEEE Robotics \& Automation Magazine},
  pp. 1--1, 02 2016.

\bibitem{Garcia1989ModelPC}
C.~E. Garcia, D.~M. Prett, and M.~Morari, ``Model predictive control: Theory
  and practice - a survey,'' \emph{Automatica}, vol.~25, pp. 335--348, 1989.

\bibitem{book}
J.~B~Rawlings and D.~Mayne, \emph{Model Predictive Control: Theory and Design},
  01 2009.

\bibitem{DBLP:journals/corr/HoE16}
\BIBentryALTinterwordspacing
J.~Ho and S.~Ermon, ``Generative adversarial imitation learning,'' \emph{CoRR},
  vol. abs/1606.03476, 2016. [Online]. Available:
  \url{http://arxiv.org/abs/1606.03476}
\BIBentrySTDinterwordspacing

\bibitem{gillula2012guaranteed}
J.~H. Gillula and C.~J. Tomlin, ``Guaranteed safe online learning via
  reachability: tracking a ground target using a quadrotor,'' in \emph{2012
  IEEE International Conference on Robotics and Automation}.\hskip 1em plus
  0.5em minus 0.4em\relax IEEE, 2012, pp. 2723--2730.

\bibitem{akametalu2014reachability}
A.~K. Akametalu, J.~F. Fisac, J.~H. Gillula, S.~Kaynama, M.~N. Zeilinger, and
  C.~J. Tomlin, ``Reachability-based safe learning with gaussian processes,''
  in \emph{53rd IEEE Conference on Decision and Control}.\hskip 1em plus 0.5em
  minus 0.4em\relax IEEE, 2014, pp. 1424--1431.

\bibitem{berkenkamp2017safe}
F.~Berkenkamp, M.~Turchetta, A.~Schoellig, and A.~Krause, ``Safe model-based
  reinforcement learning with stability guarantees,'' in \emph{Advances in
  neural information processing systems}, 2017, pp. 908--918.

\bibitem{bastani2018verifiable}
O.~Bastani, Y.~Pu, and A.~Solar-Lezama, ``Verifiable reinforcement learning via
  policy extraction,'' in \emph{Advances in Neural Information Processing
  Systems}, 2018, pp. 2494--2504.

\bibitem{alshiekh2018safe}
M.~Alshiekh, R.~Bloem, R.~Ehlers, B.~K{\"o}nighofer, S.~Niekum, and U.~Topcu,
  ``Safe reinforcement learning via shielding,'' in \emph{Thirty-Second AAAI
  Conference on Artificial Intelligence}, 2018.

\bibitem{fisac2019bridging}
J.~F. Fisac, N.~F. Lugovoy, V.~Rubies-Royo, S.~Ghosh, and C.~J. Tomlin,
  ``Bridging hamilton-jacobi safety analysis and reinforcement learning,'' in
  \emph{2019 International Conference on Robotics and Automation (ICRA)}.\hskip
  1em plus 0.5em minus 0.4em\relax IEEE, 2019, pp. 8550--8556.

\bibitem{ivanov2019verisig}
R.~Ivanov, J.~Weimer, R.~Alur, G.~J. Pappas, and I.~Lee, ``Verisig: verifying
  safety properties of hybrid systems with neural network controllers,'' in
  \emph{Proceedings of the 22nd ACM International Conference on Hybrid Systems:
  Computation and Control}.\hskip 1em plus 0.5em minus 0.4em\relax ACM, 2019,
  pp. 169--178.

\bibitem{DBLP:journals/corr/abs-1903.01287}
\BIBentryALTinterwordspacing
M.~Fazlyab, M.~Morari, and G.~J. Pappas, ``Safety verification and robustness
  analysis of neural networks via quadratic constraints and semidefinite
  programming,'' \emph{CoRR}, vol. abs/1903.01287, 2019. [Online]. Available:
  \url{https://arxiv.org/abs/1903.01287}
\BIBentrySTDinterwordspacing

\bibitem{DBLP:journals/corr/abs-1906-04893}
\BIBentryALTinterwordspacing
M.~Fazlyab, A.~Robey, H.~Hassani, M.~Morari, and G.~J. Pappas, ``Efficient and
  accurate estimation of lipschitz constants for deep neural networks,''
  \emph{CoRR}, vol. abs/1906.04893, 2019. [Online]. Available:
  \url{http://arxiv.org/abs/1906.04893}
\BIBentrySTDinterwordspacing

\bibitem{inproceedings}
D.~Tran, ``Safety verification of cyber-physical systems with reinforcement
  learning control, emsoft 2019,'' 07 2019.

\bibitem{DBLP:journals/corr/abs-1802-03557}
\BIBentryALTinterwordspacing
W.~Xiang, D.~M. Lopez, P.~Musau, and T.~T. Johnson, ``Reachable set estimation
  and verification for neural network models of nonlinear dynamic systems,''
  \emph{CoRR}, vol. abs/1802.03557, 2018. [Online]. Available:
  \url{http://arxiv.org/abs/1802.03557}
\BIBentrySTDinterwordspacing

\bibitem{DBLP:journals/corr/abs-1905-10691}
\BIBentryALTinterwordspacing
O.~Bastani, ``Safe planning via model predictive shielding,'' \emph{CoRR}, vol.
  abs/1905.10691, 2019. [Online]. Available:
  \url{http://arxiv.org/abs/1905.10691}
\BIBentrySTDinterwordspacing

\bibitem{DBLP:journals/corr/abs-1803-08552}
\BIBentryALTinterwordspacing
K.~P. Wabersich and M.~N. Zeilinger, ``Linear model predictive safety
  certification for learning-based control,'' \emph{CoRR}, vol. abs/1803.08552,
  2018. [Online]. Available: \url{http://arxiv.org/abs/1803.08552}
\BIBentrySTDinterwordspacing

\bibitem{zhang2019safe}
\BIBentryALTinterwordspacing
W.~Zhang and O.~Bastani, ``Mamps: Safe multi-agent reinforcement learning via
  model predictive shielding.'' [Online]. Available:
  \url{https://obastani.github.io/docs/mamps.pdf}
\BIBentrySTDinterwordspacing

\bibitem{doi:10.1080/00423114.2014.902537}
\BIBentryALTinterwordspacing
Y.~Gao, A.~Gray, H.~E. Tseng, and F.~Borrelli, ``A tube-based robust nonlinear
  predictive control approach to semiautonomous ground vehicles,''
  \emph{Vehicle System Dynamics}, vol.~52, no.~6, pp. 802--823, 2014. [Online].
  Available: \url{https://doi.org/10.1080/00423114.2014.902537}
\BIBentrySTDinterwordspacing

\bibitem{6160389}
S.~{Yu}, H.~{Chen}, and F.~{Allgöwer}, ``Tube mpc scheme based on robust
  control invariant set with application to lipschitz nonlinear systems,'' in
  \emph{2011 50th IEEE Conference on Decision and Control and European Control
  Conference}, Dec 2011, pp. 2650--2655.

\bibitem{allen2014machine}
R.~E. Allen, A.~A. Clark, J.~A. Starek, and M.~Pavone, ``A machine learning
  approach for real-time reachability analysis,'' in \emph{2014 IEEE/RSJ
  International Conference on Intelligent Robots and Systems}.\hskip 1em plus
  0.5em minus 0.4em\relax IEEE, 2014, pp. 2202--2208.

\bibitem{reist2016feedback}
P.~Reist, P.~Preiswerk, and R.~Tedrake, ``Feedback-motion-planning with
  simulation-based lqr-trees,'' \emph{The International Journal of Robotics
  Research}, vol.~35, no.~11, pp. 1393--1416, 2016.

\bibitem{doi:10.1002/rnc.1758}
\BIBentryALTinterwordspacing
D.~Q. Mayne, E.~C. Kerrigan, E.~J. van Wyk, and P.~Falugi, ``Tube-based robust
  nonlinear model predictive control,'' \emph{International Journal of Robust
  and Nonlinear Control}, vol.~21, no.~11, pp. 1341--1353, 2011. [Online].
  Available: \url{https://onlinelibrary.wiley.com/doi/abs/10.1002/rnc.1758}
\BIBentrySTDinterwordspacing

\bibitem{mayne2016robust}
D.~Mayne, ``Robust and stochastic model predictive control: Are we going in the
  right direction?'' \emph{Annual Reviews in Control}, vol.~41, pp. 184--192,
  2016.

\bibitem{vapnik2013nature}
V.~Vapnik, \emph{The nature of statistical learning theory}.\hskip 1em plus
  0.5em minus 0.4em\relax Springer science \& business media, 2013.

\bibitem{liebenwein2018sampling}
L.~Liebenwein, C.~Baykal, I.~Gilitschenski, S.~Karaman, and D.~Rus,
  ``Sampling-based approximation algorithms for reachability analysis with
  provable guarantees,'' in \emph{RSS}, 2018.

\bibitem{Silver:2014:DPG:3044805.3044850}
\BIBentryALTinterwordspacing
D.~Silver, G.~Lever, N.~Heess, T.~Degris, D.~Wierstra, and M.~Riedmiller,
  ``Deterministic policy gradient algorithms,'' in \emph{Proceedings of the
  31st International Conference on International Conference on Machine Learning
  - Volume 32}, ser. ICML'14.\hskip 1em plus 0.5em minus 0.4em\relax JMLR.org,
  2014, pp. I--387--I--395. [Online]. Available:
  \url{http://dl.acm.org/citation.cfm?id=3044805.3044850}
\BIBentrySTDinterwordspacing

\bibitem{DBLP:journals/corr/BrockmanCPSSTZ16}
\BIBentryALTinterwordspacing
G.~Brockman, V.~Cheung, L.~Pettersson, J.~Schneider, J.~Schulman, J.~Tang, and
  W.~Zaremba, ``Openai gym,'' \emph{CoRR}, vol. abs/1606.01540, 2016. [Online].
  Available: \url{http://arxiv.org/abs/1606.01540}
\BIBentrySTDinterwordspacing

\end{thebibliography}

\end{document}